\documentclass[aps,twocolumn,showpacs,nofootinbib]{revtex4-1}

\usepackage{amsmath,amssymb,bm}
\usepackage[dvips]{graphicx}
\usepackage{yfonts}
\newcommand{\beq}{\begin{eqnarray}}
\newcommand{\eeq}{\end{eqnarray}}
\renewcommand\d{\partial}

\begin{document}

\author{Naoki Yamamoto}
\affiliation{Department of Physics, Keio University, Yokohama 223-8522, Japan}

\title{Chiral transport of neutrinos in supernovae: \\
Neutrino-induced fluid helicity and helical plasma instability}

\begin{abstract}
Chirality of neutrinos modifies the conventional kinetic theory and hydrodynamics, 
leading to unusual chiral transport related to quantum anomalies in field theory. 
We argue that these corrections have new phenomenological consequences for hot 
and dense neutrino gases, especially in core-collapse supernovae. We find that the 
neutrino density can be converted to the fluid helicity through the chiral vortical effect. 
This fluid helicity effectively acts as a chiral chemical potential for electrons via the 
momentum exchange with neutrinos and induces a ``helical plasma instability'' that 
generates a strong helical magnetic field. This provides a new mechanism for 
converting the gravitational energy released by the core collapse to the 
electromagnetic energy and potentially explains the origin of magnetars. The other 
possible applications of the neutrino chiral transport theory are also discussed.

\end{abstract}

\maketitle 

\section{Introduction}
\label{sec:introduction}
Neutrinos play a key role in supernova explosions. When a massive star explodes, 
most of the gravitational binding energy released by the collapse of the core is 
carried away by neutrinos. In order to understand the mechanism of the supernova 
explosion and subsequent evolutions of massive stars, it is important to treat the 
neutrino transport appropriately \cite{Kotake:2005zn,Janka:2012wk};
see Refs.~\cite{Takiwaki:2013cqa,Melson:2015tia,Muller:2015dia} for recent numerical 
simulations. Nonetheless, the conventional neutrino transport theory applied so far in 
core-collapse supernovae has missed one important feature of 
neutrinos---\emph{chirality (or left-handedness) of neutrinos}. 

Recently, it has been revealed that the chirality of particles leads to dramatic 
modifications of the conventional kinetic theory (Boltzmann equation) and hydrodynamics. 
The modified kinetic theory and hydrodynamics are now called the chiral kinetic theory 
\cite{Son:2012wh,Stephanov:2012ki,Son:2012zy,Chen:2012ca,Manuel:2013zaa,Manuel:2014dza,
Chen:2014cla,Chen:2015gta} and chiral (or anomalous) hydrodynamics \cite{Son:2009tf}, 
respectively. These transport theories can reproduce the quantum anomalies in field theory 
\cite{Adler,BellJackiw}, as well as the anomalous transport phenomena, such as the 
so-called chiral magnetic effect (CME) \cite{Vilenkin:1980fu,Nielsen:1983rb,Alekseev:1998ds,Fukushima:2008xe}
and the chiral vortical effect (CVE) \cite{Vilenkin:1979ui,Son:2009tf,Landsteiner:2011cp,Landsteiner:2012kd},
which are the currents along a magnetic field and vorticity in chiral fluids.
One expects that the anomalous transport for neutrinos, which has been hitherto 
discarded, should change the evolution of the supernova explosion and the 
properties of resulting compact stars at the qualitative level.%
\footnote{Possible importance of the chirality of electrons in core-collapse 
supernovae was previously argued in Refs.~\cite{Ohnishi:2014uea,Sigl:2015xva}. 
For other applications of anomalous transport in neutron stars, see 
Refs.~\cite{Charbonneau:2009ax,Kaminski:2014jda,Shaverin:2014xya}.}

In this paper, we will focus on the hydrodynamic regime of hot and dense neutrino 
gases. (The hydrodynamic description is valid, at least, at the core of supernovae; 
see Sec.~\ref{sec:supernovae} below.) We argue that the corrections due to the 
chirality of neutrinos lead to a number of new phenomenological consequences in 
the neutrino transport theory.

We show that the neutrino density can be transmuted to the fluid helicity 
[see Eq.~(\ref{Q}) for the definition] through the CVE. Assuming that the momentum 
exchange between neutrinos and electrons is sufficiently rapid, the fluid helicity 
{\it effectively} acts as a chiral chemical potential for electrons. Then, the fluid helicity 
induces the electric current in a magnetic field, similarly to the CME, but  \emph{without} 
the chiral chemical potential for electrons itself. We call it the ``helical magnetic effect" 
(HME). When the electromagnetic fields are dynamical, such a state with nonzero 
fluid helicity becomes unstable in a way similar to the chiral plasma instability (CPI) 
\cite{Joyce:1997uy,Boyarsky:2011uy,Akamatsu:2013pjd,Akamatsu:2014yza} and 
generates a strong magnetic field with magnetic helicity. This is a new type of 
instability that originates from the chirality of neutrinos.

This provides a new mechanism for converting the gravitational energy released during 
the core collapse to electromagnetic energy by temporarily storing it as the Fermi energy 
of neutrinos and as the energy of the helical fluid motion. In particular, it may explain the 
possible origin of the gigantic magnetic fields of magnetars \cite{Magnetar} produced after 
supernova explosions. We make an order estimate of the \emph{maximum} helical magnetic 
field at the core, $B_{\rm core} \sim 10^{18}$ Gauss, when the initial neutrino chemical 
potential is $\mu_{\nu} \sim 200$ MeV. This mechanism is analogous to the one proposed in 
Ref.~\cite{Ohnishi:2014uea}, where a large chiral chemical potential for electrons is 
considered to be produced in the weak process during core collapse, which then generates 
the strong helical magnetic field by the CPI. While such a chiral chemical potential for 
electrons might be potentially damped by the effect of the electron mass \cite{Grabowska:2014efa}, 
the fluid helicity generated in the dense neutrino medium here cannot be damped 
by the fermion mass. 

We note that, compared with the conventional proposals for the origin of magnetars, 
such as the fossil field and dynamo hypothesis \cite{Harding:2006qn, Spruit:2007bt}, 
our mechanism is distinctive in that it can naturally produce the magnetic helicity (and hence, 
the linked poloidal-toroidal magnetic fields) similarly to Ref.~\cite{Ohnishi:2014uea}. 
Note also that our mechanism just relies on the dynamics within the electroweak sector and 
does not necessitate exotic hadron or quark phases inside neutron stars, including ferromagnetic 
nuclear or quark matter \cite{Brownell,Rice,Silverstein,Makishima,Tatsumi:1999ab} 
and pion domain walls \cite{Son:2007ny,Eto:2012qd}.

Our argument in this paper is schematic. However, we expect it to capture the essence 
of qualitatively new chiral effects of neutrinos disregarded so far. To what extent our new 
mechanism is efficient at the quantitative level in core-collapse supernovae should be 
checked in the future three-dimensional (3D) neutrino-radiation hydrodynamics by taking 
into account not only the source term for neutrino-matter interactions \cite{Bruenn:1985en}, 
but also the chiral effects of neutrinos appropriately.%
\footnote{Note that our argument in this paper is mostly based on the macroscopic chiral 
hydrodynamics, but the same physics should be described by the microscopic chiral 
kinetic theory. Such a kinetic description becomes necessary in the region where the matter 
density is not large enough and hydrodynamic description breaks down
(see Sec.~\ref{sec:applicability}).}
This direction is also important for the question of the supernova explosion itself, because 
the chiral effects drastically modify the evolution and structures of the fluids and 
electromagnetic fields in supernovae, as we will show in this paper. 

This paper is organized as follows. In Sec.~\ref{sec:kinetics}, we review the chiral kinetic 
theory with the stress on the relation among the chirality, topology, and Berry curvature. 
In Sec.~\ref{sec:hydro}, we summarize the basic equations and properties of the chiral 
hydrodynamics. In Sec.~\ref{sec:neutrino}, we discuss the mechanism of the neutrino-induced 
fluid helicity in the neutrino hydrodynamics. In Sec.~\ref{sec:MHD}, we consider the helical 
magnetohydrodynamics and discuss several new helical effects including the HME and 
helicity transmutation. In Sec.~\ref{sec:supernovae}, we study the chiral hydrodynamic 
effects in core-collapse supernovae and make a simple estimate for the maximum 
magnetic field generated by our mechanism. We conclude with the outlook of our work in 
Sec.~\ref{sec:outlook}.

In the following, we set $\hbar = c = e = 1$.

\section{Chiral kinetic theory}
\label{sec:kinetics}
For completeness and generality, we first review the kinetic theory (and hydrodynamics
in the next section) for charged chiral particles \cite{Son:2012wh,Stephanov:2012ki,
Son:2012zy,Chen:2012ca,Manuel:2013zaa, Manuel:2014dza,Chen:2014cla,Chen:2015gta}, 
with the emphasis on the corrections due to the chirality. The kinetic theory and 
hydrodynamics for neutral neutrinos can be obtained by simple modifications later.

\subsection{Chirality, topology, and Berry curvature}
\label{sec:topology}
We first explain why and how the chirality of fermions should lead to modifications of 
the conventional kinetic theory. Once the kinetic theory is modified, then the 
hydrodynamics must also be modified, because the latter is derived from the former by 
coarse graining. In other words, the latter is a low-energy effective theory of the 
former at long time and long distance scales much larger than the mean free time 
and mean free path.

To illustrate the point, we consider the case with $\mu \gg T$,%
\footnote{As we will see in Sec.~\ref{sec:estimate}, this condition holds approximately 
for the neutrino gas at the core of a supernova, where the neutrino chemical potential 
is $\mu_{\nu} \sim 200$ MeV and temperature is $T \sim 10$ MeV.} where the Fermi 
surface of fermions is well defined. (Here $\mu$ is the chemical potential and $T$ is 
the temperature.) We however note that the chiral kinetic theory is not limited to this 
regime; it can be generalized at high $T$ by including the contribution from antiparticles 
appropriately \cite{Manuel:2013zaa,Manuel:2014dza}.

Consider a Fermi surface of right-handed fermions. By definition, the direction of 
momentum is always the same as that of spin for right-handed particles; when the 
end point of the momentum vector of the particle covers the two-dimensional sphere 
$S^2$ (the Fermi surface) in momentum space, then that of the spin vector also 
covers a sphere $S^2$ in spin space. Hence, there is a nontrivial mapping from $S^2$ 
in momentum space to $S^2$ in spin space, whose winding number is +1. On the 
other hand, for left-handed fermions, the direction of momentum is opposite as that of 
the spin. Again, there is a nontrivial mapping from $S^2$ in momentum space to $S^2$ 
in spin space, but the winding number is $-1$ in this case.

The effects of this nontrivial topology are incorporated in the equations of motion of a 
particle (and kinetic theory) by using the notion of the Berry curvature \cite{Berry,Volovik}. 
The Berry curvature here is a fictitious ``magnetic field" in momentum space. Associated 
with the homotopy class $\pi_2(S^2)=\mathbb{Z}$ that we argued above, the Berry 
curvature is given by the field of the monopole at ${\bm p} = {\bm 0}$ as
\beq
{\bm \Omega}_{\bm p} = \pm \frac{\bm p}{2|{\bm p}|^3}\,,
\eeq
for right- and left-handed fermions, respectively. 
The integral of the Berry curvature over the Fermi surface $S^2$ is related to 
the winding number (or the monopole charge) $k = \pm 1$ as
\beq
\label{k}
k = \frac{1}{2\pi} \int_{S^2} {\bm \Omega}_{\bm p} \cdot d{\bm S}. 
\eeq

\subsection{Equations of motion and kinetic equation}
\label{sec:kinetic_equation}
The action for a chiral particle in the presence of electromagnetic fields 
${\bm E}$, ${\bm B}$ and the Berry curvature ${\bm \Omega}_{\bm p}$ 
is given by \cite{Son:2012wh,Stephanov:2012ki,Son:2012zy,Chen:2014cla}
\beq
\label{S}
S = \int [({\bm p} + {\bm A}) \cdot d{\bm x} - (\epsilon_{\bm p} + \phi)dt
-{\bm a}_{\bm p} \cdot d {\bm p}],
\eeq
where ${\bm A}$ is the vector potential and $\phi$ is the scalar potential.
Here the effect of topology in Eq.~(\ref{k}) is incorporated by the Berry 
connection ${\bm a}_{\bm p}$, which is related to the Berry curvature 
via ${\bm \Omega}_{\bm p} = {\bm \nabla}_{\bm p} \times {\bm a}_{\bm p}$.
The path-integral derivation of Eq.~(\ref{S}) can be found in 
Refs.~\cite{Stephanov:2012ki,Chen:2014cla}.

Note that the dispersion relation for chiral fermions is also modified by the
magnetic moment as \cite{Son:2012zy,Chen:2014cla,Manuel:2014dza}
\beq
\label{epsilon}
\epsilon_{\bm p} = |{\bm p}| (1 - {\bm \Omega}_{\bm p} \cdot {\bm B}).
\eeq
Here the correction concerning the Berry curvature is required by the Lorentz 
symmetry of the system \cite{Son:2012zy,Chen:2014cla}. In the absence of the 
Berry connection, the action (\ref{S}), together with the dispersion relation 
(\ref{epsilon}), reduces to the usual one which governs the dynamics of a 
(nonchiral) charged particle in electromagnetic fields.

The equations of motion for chiral particles follow from the action (\ref{S}) as
\begin{align}
\label{x_dot0}
\dot {\bm x} & = \tilde {\bm v} + \dot {\bm p} \times {\bm \Omega}_{\bm p}, 
\\
\label{p_dot0}
\dot {\bm p} & = \tilde {\bm E} + \dot {\bm x} \times {\bm B},
\end{align}
where 
\beq
\tilde {\bm v} = \frac{\d \epsilon_{\bm p}}{\d {\bm p}}\,, \qquad 
\tilde {\bm E} = {\bm E} - \frac{\d \epsilon_{\bm p}}{\d {\bm x}}\,.
\eeq
The ``anomalous velocity" in the second term of Eq.~(\ref{x_dot0}) was introduced 
earlier in the context of condensed matter physics \cite{SundaramNiu}. 
Equations (\ref{x_dot0}) and (\ref{p_dot0}) are coupled for $\dot {\bm x}$ and $\dot {\bm p}$. 
Solving Eqs.~(\ref{x_dot0}) and (\ref{p_dot0}) in terms of solely $\dot {\bm x}$ and $\dot {\bm p}$, 
one has
\begin{align}
\label{x_dot}
\sqrt{\omega}\dot {\bm x} & = \tilde {\bm v} + \tilde {\bm E} \times {\bm \Omega}_{\bm p} + 
(\tilde {\bm v} \cdot {\bm \Omega}_{\bm p}){\bm B},
\\
\label{p_dot}
\sqrt{\omega}\dot {\bm p} & = \tilde {\bm E} + \tilde {\bm v} \times {\bm B} 
+ (\tilde {\bm E} \cdot {\bm B}) {\bm \Omega}_{\bm p},
\end{align}
where $\omega = (1 + {\bm B} \cdot {\bm \Omega}_{\bm p})^2$.

We now recall the Boltzmann equation,
\beq
\label{Boltzmann}
\frac{\d n_{\bm p}}{\d t} + \dot {\bm x} \cdot \frac{\d n_{\bm p}}{\d {\bm x}}
+ \dot {\bm p} \cdot \frac{\d n_{\bm p}}{\d {\bm p}} = C[n_{\bm p}],
\eeq
where $n_{\bm p}$ is the distribution function for chiral fermions and $C[n_{\bm p}]$ is 
the collision term. Substituting Eqs.~(\ref{x_dot}) and (\ref{p_dot}) into Eq.~(\ref{Boltzmann}), 
we have the kinetic equation for chiral particles \cite{Son:2012zy,Manuel:2014dza},
\begin{align}
\label{CKT}
\frac{\d n_{\bm p}}{\d t} + \frac{1}{\sqrt{\omega}} \left(\tilde {\bm v} + \tilde {\bm E} \times {\bm \Omega}_{\bm p} 
+ (\tilde {\bm v} \cdot {\bm \Omega}_{\bm p}){\bm B} \right) \cdot \frac{\d n_{\bm p}}{\d {\bm x}}
\nonumber \\ 
+ \frac{1}{\sqrt{\omega}} \left(\tilde {\bm E} + \tilde {\bm v} \times {\bm B} 
+ (\tilde {\bm E} \cdot {\bm B}) {\bm \Omega}_{\bm p} \right) \cdot \frac{\d n_{\bm p}}{\d {\bm p}} 
= C[n_{\bm p}].
\end{align}
This is the chiral kinetic equation. Without the corrections of the Berry curvature, 
Eq.~(\ref{CKT}) reduces to the familiar kinetic equation in electromagnetic fields. 
Note that the conventional kinetic theory cannot distinguish between right- and 
left-handed particles. In the chiral kinetic theory (\ref{CKT}), on the other hand, 
they can be distinguished through the Berry curvature, because the signs of 
${\bm \Omega}_{\bm p}$ are opposite between them. 

From Eqs.~(\ref{x_dot}) and (\ref{p_dot}), one can define the particle number 
and current as \cite{Son:2012wh,Stephanov:2012ki,Son:2012zy}
\begin{align}
\label{n_CKT}
n &= \int \frac{d^3{\bm p}}{(2\pi)^3} \sqrt{\omega}n_{\bm p}\,,
\\
\label{j_CKT}
{\bm j} &= \int \frac{d^3{\bm p}}{(2\pi)^3} \sqrt{\omega} \dot {\bm x} n_{\bm p} \nonumber \\
& = \int \frac{d^3{\bm p}}{(2\pi)^3} \left(\tilde {\bm v} + \tilde {\bm E} \times {\bm \Omega}_{\bm p} 
+ (\tilde {\bm v} \cdot {\bm \Omega}_{\bm p}){\bm B} \right)n_{\bm p}\,,
\end{align}
up to the ambiguity of the shift, $\tilde {\bm j} = {\bm j} + {\bm \nabla} \times {\bm a}$ with 
any vector ${\bm a}$, which does not affect the ``continuity equation" [see Eq.~(\ref{anomaly}) 
below] because ${\bm \nabla} \cdot ({\bm \nabla} \times {\bm a}) = 0$. 
Note that the phase-space modifications in Eqs.~(\ref{n_CKT}) and (\ref{j_CKT}) 
originate from the correction concerning ${\bm \Omega}_{\bm p}$ in the equation of 
motion~(\ref{x_dot0}) \cite{Xiao:2005, Duval:2005}. The second and third terms in 
Eq.~(\ref{j_CKT}) yield the anomalous Hall effect and the CME, respectively. For the 
derivation of the energy-momentum tensor $T^{\mu \nu}$ in the chiral kinetic theory, 
see Ref.~\cite{Son:2012zy}.

In the absence of electromagnetic fields, the expression of ${\bm j}$  can be found as 
\cite{Son:2012zy,Chen:2014cla}
\beq
{\bm j} = \int \frac{d^3{\bm p}}{(2\pi)^3} \left({\bm v} - |{\bm p}|
{\bm \Omega}_{\bm p} \times {\bm \nabla} \right) n_{\bm p}\,,
\eeq
by appropriately choosing the vector ${\bm a}$ above such that the energy and 
momentum conservations are satisfied. The second term is the magnetization current 
that stems from the magnetic moment of the chiral fermion. In the hydrodynamic regime, 
it yields the CVE in a vorticity \cite{Chen:2014cla}. 

By multiplying Eq.~(\ref{CKT}) by the factor $\sqrt{\omega}$ and integrating over ${\bm p}$, 
one finds that the current conservation is modified to \cite{Son:2012wh,Stephanov:2012ki,Son:2012zy}
\begin{align}
\label{anomaly}
\d_t n + {\bm \nabla} \cdot {\bm j} &= - \int \frac{d^3{\bm p}}{(2\pi)^3}
\left({\bm \Omega}_{\bm p} \cdot \frac{\d n_{\bm p}}{\d {\bm p}} \right){\bm E} \cdot {\bm B}
\nonumber \\ 
&= \pm \frac{1}{4\pi^2} {\bm E} \cdot {\bm B}\,,
\end{align}
for right- and left-handed fermions, respectively. Here we assumed that the 
collision term satisfies
\beq
\int \frac{d^3{\bm p}}{(2\pi)^3} \sqrt{\omega}C[n_{\bm p}] = 0.
\eeq
Equation (\ref{anomaly}) is the quantum violation of the current conservation, 
known as the quantum anomaly in field theory \cite{Adler,BellJackiw}. 

The quantum anomaly, CME, and CVE above are the consequences of the chirality 
of particles, which will be important in the following discussion. For the reformulation 
of the chiral kinetic theory with collisions (but without electromagnetic fields) in a 
manifestly Lorentz-invariant manner, see Ref.~\cite{Chen:2015gta}.

\section{Chiral hydrodynamics}
\label{sec:hydro}
The chiral hydrodynamics can be obtained by taking the hydrodynamic limit of the 
chiral kinetic theory or from the underlying quantum field theory. The new corrections 
to the conventional relativistic hydrodynamics are the quantum anomaly, CME, and 
CVE, which will be expressed on the right-hand side of Eq.~(\ref{dj}), the second and 
third terms on the right-hand side of Eq.~(\ref{j}), respectively, below. These corrections 
originate from the chirality of fermions, as we have seen in Sec.~\ref{sec:kinetics}.

Originally, the corrections to the conventional relativistic hydrodynamics have been 
observed by using the gauge-gravity duality \cite{Erdmenger:2008rm,Banerjee:2008th}.
Later, the chiral hydrodynamics was derived based on the second law of thermodynamics 
without reference to the gravity \cite{Son:2009tf,Neiman:2010zi}, up to one numerical 
coefficient [the coefficient $D$ in Eqs.~(\ref{xi_B}) and (\ref{xi})]. More recently, this 
coefficient was found to be related to the mixed gauge-gravitational anomaly in field 
theory \cite{Landsteiner:2011cp,Landsteiner:2012kd,Golkar:2012kb,Jensen:2012kj}.
For other attempts to derive the chiral hydrodynamics from kinetic theory without Berry 
curvature corrections, see Refs.~\cite{Loganayagam:2012pz,Gao:2012ix}.

The relativistic hydrodynamic equations for single charged chiral fermions are given by 
the energy and momentum conservations for the energy-momentum tensor $T^{\mu \nu}$
and the anomaly relation for the electric current $j^{\mu}$ \cite{Son:2009tf},%
\footnote{Throughout the paper, we use the ``mostly minus" metric 
$g^{\mu \nu} = {\rm diag}(1,-1,-1,-1)$ unlike Ref.~\cite{Son:2009tf}.}
\begin{gather}
\label{dT}
\d_{\mu}T^{\mu\nu} = F^{\nu\lambda} j_{\lambda}, \\
\label{dj}
\d_{\mu}j^{\mu} = -C E^{\mu} B_{\mu}.
\end{gather}
Here $F^{\mu \nu}$ is the field strength, the electric and magnetic fields, 
$E^{\mu}=F^{\mu \nu} u_{\nu}$ and $B^{\mu}=\frac{1}{2}\epsilon^{\mu \nu \alpha \beta}u_{\nu}F_{\alpha \beta}$,
are defined in the fluid rest frame, and $u^{\mu}=\gamma(1, {\bm v})$ is the local fluid velocity. 
The right-hand sides of Eqs.~(\ref{dT}) and (\ref{dj}) express the work done by electromagnetic 
fields and the quantum anomaly with $C= \pm 1/(4\pi^2)$ for right- and left-handed fermions, 
respectively [see Eq.~(\ref{anomaly})].

When the electromagnetic fields are dynamical, their evolution is described 
by Maxwell's equations,
\beq
\d_{\nu} F^{\nu \mu} = j^{\mu}.
\eeq

In the Landau-Lifshitz frame \cite{Landau} where the energy diffusion is absent
and the fluid velocity is proportional to the energy current, the constitutive 
equations for $T^{\mu \nu}$ and $j^{\mu}$ are given by \cite{Son:2009tf}
\begin{gather}
\label{T}
T^{\mu \nu}=(\epsilon+P)u^{\mu}u^{\nu}-Pg^{\mu \nu}+\tau^{\mu \nu}, \\
\label{j}
j^{\mu}=nu^{\mu} + \xi_B B^{\mu} + \xi \omega^{\mu} + \nu^{\mu}.
\end{gather} 
Here $\epsilon$ is the energy density, $P$ is the pressure, $n$ is the charge density, 
and $\omega^{\mu}=\epsilon^{\mu \nu \alpha \beta} u_{\nu} \d_{\alpha} u_{\beta}$
is the vorticity. The dissipative terms, $\nu^{\mu}$ and $\tau^{\mu \nu}$, denote the 
particle diffusion and viscous stress tensor (see Refs.~\cite{Son:2009tf,Landau} 
for the detailed expressions, which are not important for our purposes).
The anomalous terms, $\xi_B$ and $\xi$, denote the transport coefficients of the 
CME \cite{Vilenkin:1980fu,Nielsen:1983rb,Alekseev:1998ds,Fukushima:2008xe} 
and CVE \cite{Vilenkin:1979ui,Son:2009tf,Landsteiner:2011cp,Landsteiner:2012kd}
and are given in this frame by \cite{Son:2009tf,Neiman:2010zi,Landsteiner:2012kd}
\begin{gather}
\label{xi_B}
\xi_B = C \mu \left(1-\frac{1}{2} \frac{n\mu}{\epsilon+P} \right)
- \frac{D}{2}\frac{nT^2}{\epsilon+P} \,, \\
\label{xi}
\xi = \frac{C}{2} \mu^2 \left(1-\frac{2}{3} \frac{n\mu}{\epsilon+P} \right) 
+ \frac{D}{2} T^2 \left(1- \frac{2 n \mu }{\epsilon+P} \right) \,.
\end{gather}
Here $D=\pm1/12$ is the coefficient of the mixed gauge-gravitational anomaly for 
right- and left-handed fermions, respectively 
\cite{Landsteiner:2011cp,Landsteiner:2012kd,Golkar:2012kb,Jensen:2012kj}.

Note that the hydrodynamic equations possess the ambiguity associated with the 
choice of the local fluid rest frame, 
and so do the expressions of the coefficients of the CME and CVE above. Here we 
choose the Landau-Lifshitz frame, so that the slow variables of the theory coincide
with the conserved quantities, $\epsilon$, $T^{0i}$, and $n$ \cite{Minami:2012hs}.

\subsection{Transport equation for ${\bm v}$}
\label{sec:v}
Let us consider the nonrelativistic limit of the bulk fluid velocity, $v \equiv |{\bm v}| \ll 1$.%
\footnote{In the core-collapse supernovae that we will discuss in Sec.~\ref{sec:supernovae}, 
the typical fluid velocity satisfies $v \ll 1$ before and after the core bounce 
\cite{Liebendoerfer:2003es,Buras:2005tb}.} In the derivative expansion of hydrodynamics, 
we further assume that $\d_t \sim {\bm v} \cdot {\bm \nabla}$. This ensures that 
$E \sim v B$, where $E \equiv |{\bm E}|$ and $B \equiv |{\bm B}|$. 

We now write down the hydrodynamic equations explicitly using 
hydrodynamic variables, ${\bm v}$, $\epsilon$, $P$, and ${\bm B}$.
The transverse component of the hydrodynamic equations with respect 
to the fluid velocity $u_{\nu}$ is obtained by multiplying by the projector, 
$P^{\rho}_{\nu} = g^{\rho}_{\nu} - u^{\rho} u_{\nu}$.
Taking the spatial component $\rho=i$, we have%
\footnote{Taking the longitudinal component of the hydrodynamic equation
and using the thermodynamic relations, 
one obtains the relation of the entropy production, 
$\d_{\mu}(s u^{\mu}) = -\sigma E^{\mu} E_{\mu} \geq 0$.}
\beq
\label{v0}
(\epsilon + P)(\d_t + {\bm v} \cdot {\bm \nabla}) {\bm v}
= - {\bm \nabla} P + {\bm j} \times {\bm B} + \nu {\bm \nabla}^2 {\bm v}.
\nonumber \\
\eeq
Here we ignored the terms ${\bm v}\d_t P$ and ${\bm v}({\bm j}\cdot {\bm E})$
on the right-hand side of Eq.~(\ref{v0}), since they are suppressed compared with 
${\bm \nabla}P$ and ${\bm j} \times {\bm B}$, respectively, 
as $v \d_t P \sim v^2 \d_s P \ll \d_s P$ and $v (j E) \sim v^2 (j B) \ll jB$
in the counting scheme above (with $\d_s$ the spatial derivative).

Equation (\ref{v0}) can be rewritten by using Maxwell's equations.
In Amp\`ere's law, the displacement current $\d_t {\bm E}$ is negligible
as $\d_t E \sim v^2 \d_s B \ll \d_s B \sim j$, and so
\beq
\label{Ampere}
{\bm j} = {\bm \nabla} \times {\bm B}. 
\eeq
Substituting it into Eq.~(\ref{v0}) and using
\beq
({\bm \nabla}  \times {\bm B}) \times {\bm B} = - \frac{1}{2} {\bm \nabla} B^2
+ (\bm B \cdot {\bm \nabla}){\bm B}, 
\eeq
we have
\begin{align}
\label{v}
(\epsilon + P)(\d_t + {\bm v} \cdot {\bm \nabla}) {\bm v}
= &- {\bm \nabla} \left(P + \frac{B^2}{2} \right) + (\bm B \cdot {\bm \nabla}){\bm B} \nonumber \\
& + \nu {\bm \nabla}^2 {\bm v}.
\end{align}
Hence, the transport equation for ${\bm v}$ is the same as the usual one \cite{Davidson}.

\subsection{Transport equation for ${\bm B}$}
\label{sec:B}
We are interested in the time scale larger than $1/\sigma$, during which charge 
diffuses immediately. In this regime, we can assume the local charge neutrality.
When $v \ll 1$, the spatial component of the electric current in Eq.~(\ref{j}) is
\beq
\label{j_e}
{\bm j}_e = \sigma ({\bm E} + {\bm v} \times {\bm B}) 
+ \xi {\bm \omega} + \xi_B {\bm B},
\eeq
where we ignored the term $\xi_B {\bm v } \times {\bm E}$, since it is suppressed 
compared with $\xi_B {\bm B}$ as $v E \sim v^2 B \ll B$. 

By eliminating ${\bm j}$ from Eqs.~(\ref{Ampere}) and (\ref{j_e}), and rewriting it in 
terms of ${\bm E}$, we have 
\beq
{\bm E} = - {\bm v } \times {\bm B} + \eta ({\bm \nabla} \times {\bm B} 
- \xi {\bm \omega} - \xi_B {\bm B}),
\eeq 
where $\eta = 1/\sigma$ is the resistivity. Substituting it into Faraday's law,
$\d_t {\bm B} = - {\bm \nabla} \times {\bm E}$, we have the transport equation that 
describes the dynamical evolution of ${\bm B}$,
\beq
\label{B}
\d_t {\bm B} = {\bm \nabla} \times ({\bm v} \times {\bm B})
+ \eta [{\bm \nabla}^2 {\bm B} + {\bm \nabla} \times 
(\xi {\bm \omega} + \xi_B {\bm B})].
\nonumber \\
\eeq

In the limit of perfect conductor, $\sigma = \infty$ (or $\eta = 0$), it reduces to 
\beq
\label{B_dissipationless}
\d_t {\bm B} = {\bm \nabla} \times ({\bm v} \times {\bm B}).
\eeq
Note that Eq.~(\ref{B_dissipationless}) is independent of $\xi$ and $\xi_B$; when 
$\eta=0$, the transport equation for ${\bm B}$ does not receive any correction in 
the presence of the CME and CVE. One consequence of this observation is that the 
conventional Alfv\'en's theorem stating that the magnetic field lines are frozen to the 
fluid motion remains applicable in chiral magnetohydrodynamics (MHD) 
for $\eta = 0$; see, e.g., Ref.~\cite{Davidson} for the proof.

\subsection{Conservation law of helicity}
\label{sec:helicity}
As we have seen above, the chiral effects do not affect the transport equations 
for ${\bm v}$ and ${\bm B}$ in the dissipationless limit. Note, however, that the 
chiral effects lead to the modification of the particle-number current, and 
consequently, to the modification of the conservation law of helicity.
We show this by using the current conservation in Eq.~(\ref{j}) and the anomaly 
relation in Eq.~(\ref{dj}) (see also Ref.~\cite{Avdoshkin:2014gpa}). Here we 
ignore the effects of dissipation, which we will briefly discuss in Sec.~\ref{sec:MHD}.

Performing the volume integral of Eq.~(\ref{dj}), we obtain
\beq
\label{conservation0}
\frac{d}{dt} \int d^3 {\bm x} \left(j^0 + \frac{C}{2} {\bm A} \cdot {\bm B} \right) = 0 \,,
\eeq
where we dropped the surface term assuming that ${\bm j}$ vanishes at infinity. 
Ignoring the $\gamma$ factor in Eq.~(\ref{j}), $j^0$ reads
\beq
\label{j0}
j^0 = n + \xi {\bm v} \cdot {\bm \omega} + \xi_B {\bm v} \cdot {\bm B}.
\eeq
Substituting Eq.~(\ref{j0}) into Eq.~(\ref{conservation0}), 
one gets the following conservation law,
\begin{gather}
\label{conservation}
\frac{d}{dt} Q_{\rm tot} = 0 \,,  \\ 
\label{Q_total}
Q_{\rm tot} \equiv Q_{\rm chi} + \frac{C}{2} Q_{\rm mag} + \xi Q_{\rm flu} + \xi_B Q_{\rm mix} \,,
\end{gather}
where
\begin{subequations}
\label{Q}
\begin{align}
Q_{\rm chi} &= \int d^3 {\bm x} \ n,\\ 
Q_{\rm mag} &= \int d^3 {\bm x} \ {\bm A} \cdot {\bm B},\\
Q_{\rm flu} &= \int d^3 {\bm x} \ {\bm v} \cdot {\bm \omega},\\
Q_{\rm mix} &= \int d^3 {\bm x} \ {\bm v } \cdot {\bm B}.
\end{align}
\end{subequations}
In plasma physics, $Q_{\rm mag}$, $Q_{\rm flu}$, and $Q_{\rm mix}$ are called 
the magnetic helicity, fluid helicity, and cross helicity, respectively \cite{Davidson}.
Equation~(\ref{conservation}) thus stands for the conservation of helicity.%
\footnote{A similar conservation law in a different frame from ours was previously 
obtained in Ref.~\cite{Avdoshkin:2014gpa}.}
Note that $Q_{\rm mag}$, $Q_{\rm flu}$, and $Q_{\rm mix}$  have different 
mass dimensions, $0$, $-2$, and $-1$, and that they are all parity-odd quantities. 

\begin{figure}[t]
\begin{center}
\includegraphics[width=6cm]{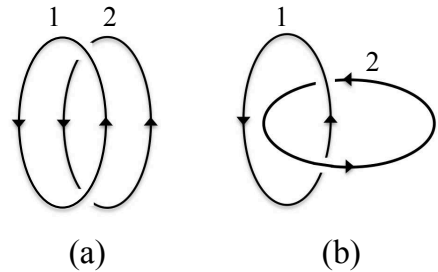}
\end{center}
\vspace{-0.7cm}
\caption{Configurations of (a) zero linking number and (b) nonzero linking number
of magnetic and vortex lines.}
\label{fig:linking}
\end{figure}

The geometric meanings of $Q_{\rm mag}$, $Q_{\rm flu}$, and $Q_{\rm mix}$ are 
well known \cite{Davidson}. The magnetic helicity $Q_{\rm mag}$ is a measure 
of linkage between two magnetic lines; $Q_{\rm mag}=0$ if the two magnetic lines 
are not linked as shown in Fig.~\ref{fig:linking}(a), while $Q_{\rm mag} \neq 0$ if 
they are linked to each other as shown in Fig.~\ref{fig:linking}(b). Similarly, the fluid 
helicity $Q_{\rm flu}$ is the degree of linkage of two vortex lines, and the cross 
helicity $Q_{\rm mix}$ is the degree of linkage of a magnetic line and a vortex line.
Recalling that the chirality of fermions is also related to the winding number 
from the momentum space to the spin space as we have seen in Sec.~\ref{sec:topology}, 
Eq.~(\ref{conservation}) may also be regarded as the ``conservation of topology."

According to the conservation law in Eq.~(\ref{conservation}), $Q_{\rm chi}$ alone is 
not conserved, but what is conserved is the combination $Q_{\rm tot}$ in Eq.~(\ref{Q_total}). 
This implies that $Q_{\rm chi}$ could be converted to $Q_{\rm mag}$, $Q_{\rm flu}$, 
and $Q_{\rm mix}$ by the hydrodynamic evolutions \cite{Avdoshkin:2014gpa}. However, 
so far, only the physical mechanism that converts $Q_{\rm chi}$ to $Q_{\rm mag}$ is known 
as the CPI \cite{Joyce:1997uy,Boyarsky:2011uy,Akamatsu:2013pjd,Akamatsu:2014yza}. 
If physical processes for other helicity conversions really exist, it follows that, even for 
neutral chiral plasmas like neutrino gases, where $Q_{\rm mag}$ and $Q_{\rm mix}$ are 
absent, $Q_{\rm chi}$ can be converted to $Q_{\rm flu}$.

In the following, we explicitly show how the transfer from $Q_{\rm chi}$ to 
$Q_{\rm flu}$ can occur in the chiral hydrodynamics for neutrinos.

\section{Neutrino hydrodynamics}
\label{sec:neutrino}
Let us now consider the hydrodynamic regime of the neutrino gas at high density 
and/or temperature. Although neutrinos interact only very weakly with other particles 
and the mean free path of neutrinos, $l_{\rm mfp}$, is quite large in typical environments, 
one can consider the dynamics of the system at the length scale $L \gg l_{\rm mfp}$. 
In this regime, the neutrino medium can be described by the neutral chiral 
hydrodynamics. As we will review in Sec.~\ref{sec:applicability}, the matter density in 
core-collapse supernovae is so high (and $l_{\rm mfp}$ becomes so small) that the 
hydrodynamic description for neutrino gases is valid for a rather small length scale 
compared with astrophysical scales. 

For the dense neutrino gas that does not couple to electromagnetic fields, 
the hydrodynamic equation (\ref{v}) is
\beq
\label{v_nu}
(\epsilon + P)(\d_t + {\bm v} \cdot {\bm \nabla}) {\bm v}
= - {\bm \nabla} P + \nu {\bm \nabla}^2 {\bm v}.
\eeq
The current conservation is given by
\beq
\label{j_nu}
\d_t (n + \xi {\bm v} \cdot {\bm \omega}) + {\bm \nabla} \cdot {\bm j} = 0, \qquad 
{\bm j} = n {\bm v} + \xi {\bm \omega},
\eeq
where $n$ is the neutrino density. 
(In this section, we suppress the index $\nu$ that stands for neutrinos for simplicity.)
Note that the hydrodynamic equation (\ref{v_nu}) is the same as the usual one for 
neutral plasmas, but the current conservation law and the current in Eq.~(\ref{j_nu}) 
have the corrections due to the CVE.

\subsection{Conservation law of helicity}
In the neutrino hydrodynamics, the conservation of helicity is obtained by eliminating 
the electromagnetic fields in Eq.~(\ref{conservation}) as
\beq
\label{Q_nu}
\frac{d}{dt} \left(Q + \xi Q_{\rm flu} \right) = 0 \,,
\eeq
where $Q$ is the total neutrino number. 
Note that the total neutrino number $Q$ itself is not conserved. 

Usually, one might expect that $Q$ must be conserved, but this is not necessarily 
the case when one takes into account the quantum effects. The well-known example 
is the baryon and lepton number violations by the quantum anomalies in the 
standard model (see, e.g., Ref.~\cite{Schwartz}). The nonconservation of $Q$ here 
is a consequence of the mixed gauge-gravitational anomaly in the hydrodynamics 
regime that appears even in a flat spacetime without gauge fields 
\cite{Landsteiner:2011cp,Landsteiner:2012kd,Golkar:2012kb,Jensen:2012kj}.

\subsection{Neutrino-induced fluid helicity (from $Q$ to $Q_{\rm flu}$)}
\label{sec:fh}
Here we illustrate how the fluid helicity can be generated in the neutrino chiral 
hydrodynamics in a simple setup. With keeping the application to neutrino gases in 
core-collapse supernovae in mind, we assume that $\mu \gg T$.%
\footnote{This assumption is not essential in the following argument, 
but just simplifies the expression of the transport coefficient $\xi$ in Eq.~(\ref{xi}).} 
Using the thermodynamic relation, $\epsilon + P = \mu n + Ts \simeq \mu n$, 
the transport coefficient $\xi$ in Eq.~(\ref{xi}) reduces to
\beq
\label{xi_nu}
\xi \simeq -\frac{\mu^2}{24 \pi^2} \,,
\eeq 
where we used $C=-1/(4\pi^2)$ for left-handed neutrinos.

\begin{figure}[t]
\begin{center}
\includegraphics[width=7cm]{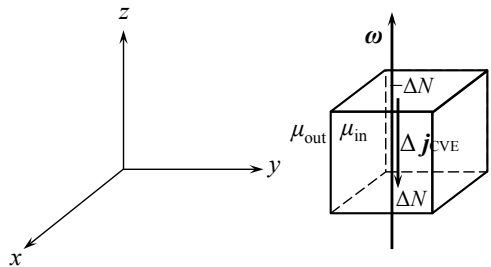}
\end{center}
\vspace{-0.7cm}
\caption{Generation of $Q_{\rm flu}$ in the neutrino hydrodynamics in a vorticity 
${\bm \omega}$. See the text for further detail.}
\label{fig:fh}
\end{figure}

As shown in Fig.~\ref{fig:fh}, we consider an infinitesimal cubic volume element 
$\Delta V= \Delta x \Delta y \Delta z$ in a vorticity in the positive $z$ direction, 
${\bm \omega} = \omega \hat {\bm z}$. 
(Note that there is no fluid velocity ${\bm v}$ in the $z$ direction in this volume element 
at the beginning.) In general, the neutrino chemical potential depends on the position, 
$\mu({\bm x})$, and the chemical potential inside this volume element can be slightly 
different from that outside this region. We denote this difference by 
$\Delta \mu \equiv \mu_{\rm in} - \mu_{\rm out}$.
Assuming $\Delta \mu >0$, the magnitude of the transport coefficient $\xi$ in 
Eq.~(\ref{xi_nu}) is larger inside the volume element by
\beq
|\Delta \xi| \simeq \frac{\mu}{12 \pi^2}\Delta \mu \,,
\eeq
to the order of $O(\Delta \mu)$, where $\mu \equiv \mu_{\rm out}$.
Because of the difference $\Delta \xi$ between the inside and the outside, 
there is an additional chiral vortical current in the negative $z$ direction, 
\beq
\Delta j^z_{\rm CVE} = -\frac{\mu \omega}{12 \pi^2} \Delta \mu \,.
\eeq 

After an infinitesimally small time interval $\Delta t$, this additional current leads to 
the accumulation of neutrinos in the lower plane and the depletion in the upper plane, 
whose number is expressed by $\Delta N = |\Delta j^z_{\rm CVE}| \Delta S \Delta t$
with $\Delta S =  \Delta x \Delta y$.
The difference of neutrino numbers between the upper and lower planes
generates the gradient of the pressure, $\d_z P(z) < 0$, in the $z$ direction
in this volume element. From the hydrodynamic equation (\ref{v_nu}), 
this gives rise to the local fluid velocity in the positive $z$ direction,
\beq
\label{v_z}
\Delta v_z = - \frac{\d_z P}{\epsilon + P} \Delta t \,,
\eeq
to the order of $O(v)$.
As a result of this process, the nonzero fluid helicity is generated in this volume element as
\beq
\Delta Q_{\rm flu} = \omega \Delta v_z \Delta V.
\eeq

The variations of the neutrino density and/or temperature and the local vorticity can occur 
in every part of the system, and generates a global fluid helicity $Q_{\rm flu}$ in total. 
Then, according to Eq.~(\ref{Q_nu}), the total neutrino number $Q$ also changes so that the 
total helicity of the system is conserved. Note here that the parity-violating transport (CVE) is 
essential to generate the parity-odd $Q_{\rm flu}$%
\footnote{We note that the configuration in Fig.~\ref{fig:fh} does not break parity in the sense 
that $\int d^3 {\bm x} \ {\bm \omega} \cdot {\bm \nabla} \mu = 0$. If this parity-odd quantity was 
nonzero, $Q_{\rm flu}$ could be generated even without the CVE. The point of our argument 
here is that $Q_{\rm flu}$ can be generated through the CVE although the configuration 
respects parity.}; this process is specific for the neutrino chiral hydrodynamics, and it does not 
occur in the usual hydrodynamics.

\section{Helical magnetohydrodynamics}
\label{sec:MHD}
We now consider a plasma which includes not only hot and/or dense neutrinos, but also 
charged electrons. We assume that the momentum exchange between neutrinos and 
electrons is sufficiently rapid, so that the fluid dynamics is described by the single fluid 
velocity ${\bm v}$ in the hydrodynamic regime of the system. Since the fluid helicity can 
be generated through the CVE for neutrinos as we have seen above, the system of 
interest is described by the MHD with finite fluid helicity---\emph{helical magnetohydrodynamics}.

For simplicity, we assume that the chiral chemical potential for electrons, 
$\mu_5 \equiv ({\mu_{\rm R}}-{\mu_{\rm L}})/2$, is much smaller than the vector chemical 
potential, $\mu_e \equiv ({\mu_{\rm R}}+{\mu_{\rm L}})/2$. 
(The extension of the following discussion to large $\mu_5$ should be straightforward.)

Taking the summation and subtraction of Eq.~(\ref{conservation}) for right- and left-handed 
fermions (electrons and neutrinos), we obtain the following relations concerning 
$Q \equiv Q_e^{\rm R} + Q_e^{\rm L} + Q_{\nu}^{\rm L}$ and 
$Q_{5} \equiv Q_e^{\rm R} - Q_e^{\rm L} - Q_{\nu}^{\rm L}$:
\begin{align}
\label{relation_Q}
\frac{d}{dt}Q &= -\frac{d}{dt}(\xi_{\nu} Q_{\rm flu})\,, 
\\
\label{relation_Q5}
\frac{d}{dt}\tilde Q_5 &= 0 \,, 
\end{align}
where 
\beq
\label{tilde_Q5}
\tilde Q_5 \equiv Q_5 + \frac{1}{4\pi^2}Q_{\rm mag} 
+ (\xi_e - \xi_{\nu}) Q_{\rm flu} 
+ \xi_e^B Q_{\rm mix}\,,
\eeq
with 
\begin{gather}
\xi_e  = \frac{\mu_e^2}{4\pi^2}\left(1 - \frac{1}{3}\frac{n_e \mu_e}{\epsilon + P} \right) 
+ \frac{T^2}{12}\left(1 - \frac{n_e \mu_e}{\epsilon + P} \right) \,,\\
\xi_{\nu}  = -\frac{\mu_{\nu}^2}{8\pi^2}\left(1 - \frac{2}{3}\frac{n_{\nu} \mu_{\nu}}{\epsilon + P} \right) 
- \frac{T^2}{24}\left(1 - \frac{2 n_{\nu} \mu_{\nu}}{\epsilon + P} \right) \,,\\
\xi_e^B = \frac{\mu_e}{2\pi^2}\left(1 - \frac{1}{4}\frac{n_e \mu_e}{\epsilon + P} \right) 
- \frac{1}{24}\frac{n_e T^2}{\epsilon + P} \,.
\end{gather}
Equations (\ref{relation_Q}) and (\ref{relation_Q5}) represent the violation of the 
lepton number and the conservation of total helicity, respectively. Both the violation 
of the lepton number and the modifications of the axial charge originate from the 
quantum anomalies in the hydrodynamic regime. 
Note that, even for finite $\nu$ and $\eta$, the contribution of the dissipative terms 
to these relations is negligibly small when $\mu_5 \ll \mu_e, T$.

\subsection{Helical magnetic effect}
\label{sec:HME}
When the nonzero global fluid helicity $Q_{\rm flu}$ is generated from the neutrino 
medium, it necessarily implies a nonvanishing local fluid helicity 
$n_{\rm flu}({\bm x}) \equiv {\bm v} \cdot {\bm \omega}$ in some region.
Then, $n_{\rm flu}$ {\it effectively} acts as a chiral chemical potential $\mu_5$ for 
the other charged particles. In fact, $n_{\rm flu}$ has the same mass dimension 
and the same quantum numbers as $\mu_5$, and the following current as a 
response to $n_{\rm flu}$ and ${\bm B}$ is generally allowed to emerge in terms 
of ${\cal C}$, ${\cal P}$, and ${\cal T}$ symmetries:
\beq
\label{HME}
{\bm j}_{\rm HME} = \kappa_B n_{\rm flu} {\bm B}.
\eeq
Compared with the expression of the CME for Dirac particles, 
${\bm j_{\rm CME}} \propto \mu_5 {\bm B}$ 
\cite{Vilenkin:1980fu,Nielsen:1983rb,Alekseev:1998ds,Fukushima:2008xe}, the coefficient 
$\mu_5$ is replaced by $n_{\rm flu}$ with some proportionality constant $\kappa_B$ 
of order $1$. We refer to it as the ``helical magnetic effect" (HME). 

A similar effect is known as the $\alpha$-effect in plasma physics \cite{Davidson}, 
where $\langle n_{\rm flu}({\bm x}) \rangle$ with the mean value averaged over 
the turbulent fluctuations is considered to arise by helical turbulence. However, it is 
not clear, as a matter of principle, how a parity-odd fluid helicity can be created 
\emph{globally} (i.e., $Q_{\rm flu} = \int d^3 {\bm x} \langle n_{\rm flu} \rangle \neq 0$)
in the evolutions of conventional parity-preserving hydrodynamic equations.
In contrast, in our argument, the parity symmetry is broken by the chirality of 
neutrinos and the global fluid helicity can be generated naturally.

\subsection{Helical plasma instability (from $Q_{\rm flu}$ to $Q_{\rm mag}$)}
\label{sec:HPI}
When electromagnetic fields evolve dynamically in the presence of the HME 
in Eq.~(\ref{HME}), an instability similar to the CPI develops. (For the physical 
picture of the CPI, see Ref.~\cite{Akamatsu:2014yza}.) We will call it the 
``helical plasma instability" (HPI). Here we provide a derivation of the HPI in an 
analytically tractable setup.

We consider the region where $n_{\rm flu}({\bm x}) \neq 0$ and $n_{\rm mix}({\bm x}) = 0$,
and assume that the variation of the fluid helicity is much smaller than its magnitude, 
$|{\bm \nabla}n_{\rm flu}/n_{\rm flu}| \ll |n_{\rm flu}|$.
The equation for ${\bm B}$ is obtained by replacing the chiral magnetic current 
$\xi_B {\bm B}$ by the helical magnetic current $\kappa_B n_{\rm flu} {\bm B}$ 
and by turning off the chiral vortical current $\xi {\bm \omega}$ in Eq.~(\ref{B}) as
\beq
\label{B_HME}
\d_t {\bm B} = {\bm \nabla} \times ({\bm v} \times {\bm B})
+ \eta ({\bm \nabla}^2 {\bm B} + \kappa_B n_{\rm flu} {\bm \nabla} \times {\bm B}).
\eeq
Here we concentrated on the magnetic field with the momentum $k \sim n_{\rm flu}$, and 
dropped the term ${\bm \nabla} n_{\rm flu} \times {\bm B}$.
As we will show in Eq.~(\ref{k_inst}) below, the typical momentum scale of the HPI is 
actually the magnitude of the local fluid helicity $n_{\rm flu}$.

Equation (\ref{B_HME}) has unstable modes. To see this, we take $v_x=v_y=0$, 
$v_z \neq 0$ with $\d_y v_x = -\d_x v_y \neq 0$ (and we set all the other components zero)
so that $n_{\rm flu} = v_z \omega_z \neq 0$ in this region. 
We consider a perturbation of the magnetic field of the form,
\beq
{\bm B}_{\pm} = ({\bm e}_x \pm i {\bm e}_y) e^{-i \omega t + i k z},
\eeq
where the subscript $\pm$ denotes the state with helicity $\pm 1$ for 
positive $k$. Substituting it into Eq.~(\ref{B_HME}), we get the dispersion relation,
\beq
\label{dispersion}
\omega = \mp \d_y v_x + k v_z - i \eta k (k \mp \kappa_B n_{\rm flu}).
\eeq

The imaginary part of $\omega$ becomes maximal when 
\beq
\label{k_inst}
k_{\rm inst} = \pm \kappa_B n_{\rm flu}/2, 
\eeq
for which $\omega$ has the positive imaginary part.
At $k=k_{\rm inst}$, the magnetic field grows exponentially (at least initially) as
\beq
|B(t)| = |B(0)| e^{\eta (\kappa_B n_{\rm flu})^2 t/4}.
\eeq
This is the HPI. The resulting magnetic field acquires a nonzero magnetic helicity, 
similarly to the CPI \cite{Joyce:1997uy,Boyarsky:2011uy,Akamatsu:2013pjd,Akamatsu:2014yza}. 
In the present case, however, the change of the magnetic helicity is accompanied by the 
change of the fluid helicity due to the conservation of total helicity [see Eqs.~(\ref{relation_Q5}) 
and (\ref{tilde_Q5})], but not that of $\mu_5$ as in 
Refs.~\cite{Joyce:1997uy,Boyarsky:2011uy,Akamatsu:2013pjd,Akamatsu:2014yza}.

\subsection{Helical vortical effect}
\label{sec:HVE}
As we will see in Sec.~\ref{sec:cross}, the fluid helicity $Q_{\rm flu}$ can also be 
converted to the cross helicity $Q_{\rm mix}$ by the helical hydrodynamic evolution.
Here, notice that the local cross helicity $n_{\rm mix} = {\bm v} \cdot {\bm B}$
has the same mass dimension and the same quantum numbers as $\mu_e \mu_5$ 
(but not $\mu_5$ itself). Hence, similarly to the CVE for Dirac particles, 
${\bm j}_{\rm CVE} \propto \mu_e \mu_5 {\bm \omega}$ 
\cite{Vilenkin:1979ui,Son:2009tf,Landsteiner:2011cp}, $n_{\rm mix}$ can induce a 
current in a vorticity,
\beq
\label{HVE}
{\bm j}_{\rm HVE} = \kappa n_{\rm mix} {\bm \omega},
\eeq
with some constant $\kappa$ of order $1$. We call it as the ``helical vortical effect" (HVE).

One might expect that ${\bm j} \sim n_{\rm mag} {\bm B}$, with 
$n_{\rm mag} = {\bm A} \cdot {\bm B}$ being the local ``magnetic helicity," should exist as well, 
since this relation is consistent with ${\cal C}$, ${\cal P}$, and ${\cal T}$ symmetries. 
However, $n_{\rm mag}$ is not gauge invariant and does not make sense locally, unlike the 
global magnetic helicity $Q_{\rm mag}$ which is gauge invariant under the appropriate 
boundary conditions (e.g., ${\bm B} \rightarrow {\bm 0}$ at infinity). 
Therefore, we do not consider the current of the form ${\bm j} \sim n_{\rm mag} {\bm B}$. 

\subsection{From $Q_{\rm flu}$ to $Q_{\rm mix}$ and vice versa}
\label{sec:cross}
We now show, from the argument parallel to the one in Sec.~\ref{sec:fh}, that the 
cross helicity $Q_{\rm mix}$ can be generated from the fluid helicity $Q_{\rm flu}$ 
through the HME, and that $Q_{\rm flu}$ can be generated from $Q_{\rm mix}$ 
through the HVE. For the later purpose, we set $\alpha \equiv \kappa_B n_{\rm flu}$ 
and $\beta \equiv \kappa n_{\rm mix}$.

\begin{figure}[t]
\begin{center}
\includegraphics[width=3.3cm]{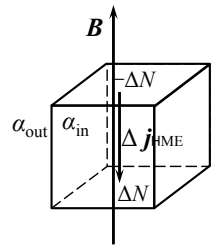}
\end{center}
\vspace{-0.5cm}
\caption{Generation of $Q_{\rm mix}$ from $Q_{\rm flu}$ in the MHD. 
See the text for further detail.}
\label{fig:cross}
\end{figure}

We consider an infinitesimal cubic volume element $\Delta V$ in a magnetic field, 
${\bm B} = B \hat {\bm z}$, as shown in Fig.~\ref{fig:cross}.  We assume that $\alpha_{\rm in}$ 
inside this volume element is smaller than $\alpha_{\rm out}$ outside the region, and we 
denote this difference by $\Delta \alpha \equiv \alpha_{\rm in} - \alpha_{\rm out}<0$. 
This difference leads to an additional helical magnetic current in the negative $z$ direction, 
\beq
\Delta j^z_{\rm HME} = B \Delta \alpha.
\eeq 
After a time $\Delta t$, this additional current leads to the increase of the charged 
particles in the lower plane by $\Delta N = \Delta j^z_{\rm HME} \Delta S \Delta t$
and the decrease in the upper plane by $\Delta N$.
Analogously to Eq.~(\ref{v_z}), this yields $v_z > 0$, and consequently, 
the nonzero cross helicity is generated in this volume element as
\beq
\Delta Q_{\rm mix} = B \Delta v_z \Delta V.
\eeq
Note that the presence of $Q_{\rm flu}$ is essential in this process to 
produce $Q_{\rm mix}$.

The generation of $Q_{\rm flu}$ from $Q_{\rm mix}$ can be understood
in a similar manner. With $\Delta \beta \equiv \beta_{\rm in} - \beta_{\rm out}<0$ 
in a vorticity ${\bm \omega} = \omega \hat {\bm z}$, we have the additional helical cortical 
current in the volume element,
\beq
\Delta j^z_{\rm HVE} = \omega \Delta \beta.
\eeq
This current leads to $v_z > 0$, which then provides the nonzero fluid helicity,
\beq
\Delta Q_{\rm flu} = \omega \Delta v_z \Delta V.
\eeq

\subsection{Towards saturation}
\label{sec:saturation}
We have seen above that different types of helicity can be converted to each other by the 
evolutions in helical MHD: conversion from $Q_{\rm flu}$ to $Q_{\rm mix}$ and $Q_{\rm mag}$,
and conversion from $Q_{\rm mix}$ to $Q_{\rm flu}$ and $Q_{\rm mag}$. 
(Here conversion from $Q_{\rm mix}$ to $Q_{\rm mag}$ occurs, at least, by combining two
processes from $Q_{\rm mix}$ to $Q_{\rm flu}$ and from $Q_{\rm flu}$ to $Q_{\rm mag}$.) 

We note that our argument here provides an explicit realization of the possible 
new ``instability" suggested in Ref.~\cite{Avdoshkin:2014gpa}. This is a kind of extension 
of the CPI in kinetic theory \cite{Akamatsu:2013pjd} to the hydrodynamic regime, where 
chiral charge of fermions is transferred not only to $Q_{\rm mag}$, but also to 
$Q_{\rm flu}$ and $Q_{\rm mix}$. As far as we know, however, we do not find any 
exponentially growing mode concerning the helicity conversion from the chiral charge to 
$Q_{\rm flu}$ and $Q_{\rm mix}$ unlike 
the CPI \cite{Joyce:1997uy,Boyarsky:2011uy,Akamatsu:2013pjd,Akamatsu:2014yza} or HPI. 
So we regard this mechanism as a ``helicity transmutation" rather than the ``instability" 
at this moment.

How the system saturates after the helicity transmutation depends on the details of 
the system, and its quantitative understanding requires 3D helical MHD simulations 
with the appropriate initial conditions. It is plausible to expect that, if the conversion 
efficiency of the helicity is sufficient, the magnitude of each term in Eq.~(\ref{tilde_Q5}) 
should be of the same order after the saturation.

\section{Application to core-collapse supernovae}
\label{sec:supernovae}
We now apply our mechanism above to core-collapse supernovae, where left-handed 
neutrinos are abundantly produced by the electron capture $p + e^- \rightarrow n + \nu_e$ 
and form a Fermi degenerate matter \cite{Kotake:2005zn,Shapiro}. The key point that has
not been appreciated so far, to our knowledge, is that the neutrino matter here is a 
\emph{chiral liquid}. Below we will concentrate on electron neutrinos (which we will denote as $\nu$), 
since only they are numerously generated in this process during the core collapse. To 
illustrate the new chiral effects of neutrinos, we ignore the general relativistic corrections.

\subsection{Applicability of neutrino hydrodynamics}
\label{sec:applicability}
Let us first discuss the applicability of the neutrino hydrodynamics at the core of the supernova. 
For this purpose, recall the expression for the mean free path of neutrinos due to the 
coherent scattering with nuclei \cite{Kotake:2005zn},
\beq
\label{mfp}
l_{\rm mfp} \sim 10^7 \ {\rm cm} \left(\frac{\rho}{10^{10} \ {\rm g}/{\rm cm}^3} \right)^{\! -\frac{5}{3}}
\left(\frac{A}{56} \right)^{\! -1} \left(\frac{Y_e}{26/56} \right)^{\! -\frac{2}{3}}\,,
\nonumber \\
\eeq
where $\rho$ is the mass density of nuclear matter, $A$ is the atomic mass number, and 
$Y_e$ is the electron fraction. Equation (\ref{mfp}) can be understood from the expression
of the mean free path, $l_{\rm mfp} = (\sigma_A n_A)^{-1}$, where $\sigma_A$ is the 
cross section and $n_A$ is the number density of nuclei. By using 
$\sigma_A \sim G_{\rm F}^2 E_{\nu}^2 A^2$ \cite{Tubbs:1975jx} and 
$n_A = \rho/(A m_N)$, where $G_{\rm F}$ is the Fermi constant, 
$E_{\nu} \simeq \mu_{e} = (3\pi^2 \rho Y_e/m_N)^{1/3}$ is the neutrino energy, and 
$m_N$ is the nucleon mass, one arrives at Eq.~(\ref{mfp}).

Substituting the typical magnitudes of the quantities appearing in Eq.~(\ref{mfp}) at the core,
$\rho \gtrsim 10^{13} \ {\rm g}/{\rm cm}^3$, $A \simeq 56$, and $Y_e \sim 0.1$, 
we have $l_{\rm mfp} \lesssim 1 \ {\rm m}$. (For the highest density 
$\rho \sim 10^{15} \ {\rm g}/{\rm cm}^3$, we have even $l_{\rm mfp} \sim 1 \ {\rm cm}$.)
Therefore, the hydrodynamic description for dense neutrino gases should be valid at the 
astrophysical length scale $L \gg l_{\rm mfp}$, at least at the core of the supernova. 
(Recall that the typical radius of the core is of order $100$ km.) In the lower density 
region where hydrodynamics for neutrinos becomes invalid, one needs to use the 
chiral kinetic theory for neutrino gases instead (see Sec.~\ref{sec:kinetics}), which 
is beyond the scope of the present paper.

\subsection{Estimate of magnetic fields}
\label{sec:estimate}
From now on, we provide a simple estimate of the maximum magnetic field that can be 
generated by our mechanism. Our estimate below should be regarded as schematic. 

The highest temperature is $T \sim 10$ MeV and the maximum neutrino and electron 
chemical potentials at the core are on the order of the nuclear scale, 
$\mu_{\nu} \sim \mu_e \sim \Lambda$, where $\Lambda \equiv 200$ MeV.%
\footnote{The numerical results of $T$ and $\mu_{\nu}$ in the neutrino-radiation 
hydrodynamics can be found, e.g., in Ref.~\cite{Roberts:2012zza}.} 
The latter fact may be understood from the near $\beta$ equilibrium condition, charge 
neutrality, and the typical lepton fraction: $\mu_{n} + \mu_{\nu} \simeq \mu_{p} + \mu_e$, 
$n_p = n_e$, and $Y_{l} \equiv (n_e + n_{\nu})/(n_n + n_p) \sim 0.1$.
Here $\mu_n$ and $\mu_p$ are the neutron and proton chemical potentials, and $n_n$ 
and $n_p$ are the neutron and proton densities, all of which are set by the nuclear scale.

As the time scale of the core collapse, $t_{\rm coll} \sim 1$ s, is much larger than the 
time scale corresponding to the typical energy scale in the chiral hydrodynamics,
$\Lambda^{-1} \sim 10^{-23}$ s, there must be sufficient time for the saturation of 
the helicity transmutation to be achieved. Hence, assuming the sufficient conversion 
efficiency of the helicity, we expect that (see Sec.~\ref{sec:saturation})
\beq
Q_{\rm mag} \sim \mu_{\nu}^2 Q_{\rm flu} \sim Q_{\nu}^0 \sim V \Lambda^3,
\eeq
after the saturation for $\mu_{\nu} \gg T$, where $Q_{\nu}^0$ is the initial neutrino number 
at the core and $V$ is the volume of the core. Then, the typical magnitude of the local 
fluid helicity $n_{\rm flu} \sim Q_{\rm flu}/V$ is of order $\Lambda$, and so is the typical 
momentum scale of the HPI from Eq.~(\ref{k_inst}), $k_{\rm inst} \sim \Lambda$. 
Considering $Q_{\rm mag} \sim V B_{\rm core}^2/k_{\rm inst} $, we obtain
\beq
B_{\rm core} \sim \Lambda^2 \sim 10^{18} \ {\rm Gauss}.
\eeq
Hence, the maximum magnetic field generated by this mechanism is set 
by the nuclear scale.

\subsection{Discussion}
\label{sec:discussion}
Finally, we discuss several important effects concerning our mechanism that we have 
ignored so far. 

\subsubsection{Helical turbulence and inverse cascade}
At first sight, the typical length scale of the magnetic field, 
$k_{\rm inst}^{-1} \sim \Lambda^{-1} \sim 10^{-15} \ {\rm m}$, generated just after the HPI 
is too small compared with astrophysical scales. Yet, there exists a mechanism, called the 
inverse cascade, that enhances a wavelength of a magnetic field to a macroscopic length 
scale by a MHD turbulence \cite{Biskamp}. The inverse cascade is known to take place 
in charged plasmas with nonzero magnetic helicity \cite{Biskamp}, which is the case for 
electromagnetic plasmas coupled with neutrinos in Sec.~\ref{sec:MHD}.%
\footnote{Precisely speaking, our situation is not completely the same as the usual charged 
plasmas with magnetic helicity \cite{Biskamp} in that the magnetic helicity is not conserved 
due to the helicity transmutation. Whether the plasma here really exhibits the inverse cascade 
or not should be clarified in the future \cite{cascade}. We note that, without the fluid velocity 
(${\bm v} = {\bm 0}$), the inverse cascade was numerically confirmed in the 
Maxwell-Chern-Simons theory \cite{Hirono:2015rla}.}  
The possible inverse cascade in the 3D helical MHD also gives rise to a large-scale coherent 
fluid motion, which is favorable for the supernova explosion itself. This should be contrasted 
with the direct cascade observed in the conventional 3D neutrino-radiation hydrodynamics 
that tends to make the explosion rather difficult \cite{Hanke:2011jf}. To what extent the inverse 
cascade is efficient in the present case should be checked in the future 3D helical MHD 
simulations.

\subsubsection{Chirality flipping}
We then consider the effects of chirality flipping by fermion masses. Since chirality 
flipping from left-handed neutrinos to right-handed ones is negligible, the generation of 
fluid helicity in the neutrino hydrodynamics in Sec.~\ref{sec:fh} should not be prevented 
by the fermion mass.%
\footnote{This should be contrasted with the case of chiral electrons produced in the weak 
process during core-collapse supernovae \cite{Ohnishi:2014uea}, where chirality flipping 
by the electron mass might damp the CPI \cite{Grabowska:2014efa}.}
On the other hand, chirality flipping by the electron mass can decrease the chiral charge 
of electrons, $Q_5$, even if it is generated in the weak process \cite{Ohnishi:2014uea} or 
by the inverse process from $Q_{\rm mag}$, $Q_{\rm flu}$, or $Q_{\rm mix}$. Although 
such an inverse process is generally allowed, we expect it to cease after the saturation of 
the helicity transmutation, as we argued in Sec.~\ref{sec:saturation}. 
Then, $Q_{\rm mag}$, $Q_{\rm flu}$, and $Q_{\rm mix}$, which are not affected by the nonzero 
electron mass, remain nonzero after the saturation, while $Q_5$ is damped. The resulting 
$Q_{\rm mag}$, $Q_{\rm flu}$, and $Q_{\rm mix}$ can play important roles in the subsequent 
evolutions of the stars and, especially, $Q_{\rm mag}$ possibly accounts for the origin of 
magnetars. Whether this expectation is correct or not needs to be tested in the helical MHD 
simulations as well.

\subsubsection{Profiles of proto-neutron stars}
One might wonder if all the newly born neutron stars just after supernova explosions become 
magnetars by this mechanism. Note that our estimate here provides a \emph{maximum} 
magnetic field assuming the sufficient conversion efficiency of the helicity of neutrinos into 
that of electromagnetic fields. To clarify this point, it is also important to understand the 
realistic conversion efficiency that should depend on individual profiles of proto-neutron stars,
such as the initial rotation and magnetic field.

\section{Outlook}
\label{sec:outlook}
In this paper, we pointed out that the chirality (or the left-handedness) of neutrinos leads to 
a number of new phenomenological consequences in a hot and dense neutrino medium 
and charged plasmas coupled with it. In particular, we found that the neutrino density can 
be converted to the fluid helicity of the neutrino medium. The resulting fluid helicity 
effectively acts as a chiral chemical potential for other charged particles, and induces 
various helical effects, such as the helical magnetic effect, helical vortical effect, and 
helical plasma instability. Through these helical effects, the fluid helicity can also be 
converted to the magnetic helicity and cross helicity. 

In the context of core-collapse supernovae, this provides a new mechanism for converting 
the gravitational energy released by the core collapse to the fluid energy and the 
electromagnetic energy, which may explain the possible origin of magnetars. 
Since our mechanism modifies the structure and evolution of the fluid motion and those of 
electromagnetic fields, this is relevant to the question of the supernova explosion itself.

There are several future directions of our work. Among others, our new mechanism should 
be numerically checked in the 3D \emph{chiral} neutrino-radiation hydrodynamics. It would 
be important to see if and how the conventional picture of the supernova explosion is 
modified quantitatively or even qualitatively. It would also be interesting to study the possible 
impacts of new collective modes specific for neutrino gases due to the chiral effects, 
such as the analogue of the chiral Alfv\'en wave in a rotation \cite{Yamamoto:2015ria}, 
chiral vortical wave \cite{Jiang:2015cva}, and chiral heat wave \cite{Chernodub:2015gxa}.

Our primary interest in this paper has been focused on the dense neutrino medium in 
core-collapse supernovae. However, the mechanism of the helicity transmutation itself is 
general and is applicable to other chiral matter as well, such as the electroweak plasma 
in the early Universe \cite{Joyce:1997uy,Boyarsky:2011uy}, quark-gluon plasmas created 
in heavy ion collisions \cite{Kharzeev:2007jp,Fukushima:2008xe} (see also 
Ref.~\cite{Kharzeev:2015znc} for a recent review), and Weyl semimetals 
\cite{Vishwanath,BurkovBalents,Xu-chern}. One may also consider the possible generation 
of lepton number from the fluid helicity in the cosmology. We defer these questions to future work.

\acknowledgments 
The author thanks Y.~Akamatsu, A.~Ohnishi, and T.~Takiwaki for useful discussions 
and correspondence. This work was supported, in part, by JSPS KAKENHI Grants 
No.~26887032 and MEXT-Supported Program for the Strategic Research Foundation
at Private Universities, ``Topological Science'' (Grant No.~S1511006).

\end{document}